\documentclass[a4paper,12pt,fleqn]{article}       
\usepackage[doublespacing]{setspace}            

\usepackage[english]{babel}
\usepackage{newtxtext}
\usepackage{mathtools}
\usepackage{amsthm}
\usepackage{newtxmath}
\usepackage{graphicx}
\usepackage{cite}

\usepackage{pfr_preprint_2026_p}

\usepackage{bm}

\newcommand{\p}{\partial}

\newcommand{\eq}{{\mathrm{eq}}}
\newcommand{\pt}{{\mathrm{pt}}}
\newcommand{\yhat}{\hat{\bm{y}}}
\newcommand{\zhat}{\hat{\bm{z}}}

\begin{document}

\title{Plasmoid Instability in Inhomogeneous Current Sheets}

\author{Kazuma~IKEDA and Ryusuke~NUMATA}

\affiliation{
    Graduate School of Information Science, University of Hyogo, Kobe 650-0047, Japan}

\email{ryusuke.numata@gmail.com}

\begin{abstract}
Plasmoid instability in inhomogeneous current sheets is investigated using electromagnetic gyrokinetic simulations of collisionless magnetic reconnection. Simulations with and without small random perturbations are compared to examine the effects of symmetry breaking and spatial inhomogeneity. The current-sheet width immediately before disruption remains close to the electron skin depth, whereas the disruption aspect ratio varies significantly between primary and later current sheets. Random perturbations also produce asymmetric plasmoid topologies in which neighboring X points no longer share a common separatrix. These results suggest that spatial inhomogeneity is an important factor in the disruption of dynamically evolving current sheets.
\end{abstract}



\maketitle  


\section{Introduction}
\label{sec:intro}

Magnetic reconnection is a fundamental process commonly observed in astrophysical and laboratory plasmas. It changes the magnetic-field topology and rapidly converts the free energy stored in the magnetic field into plasma kinetic and thermal energy and, in some cases, non-thermal particle energy.

A current sheet plays a central role in magnetic reconnection. The tearing instability of a current-sheet configuration was formulated in resistive magnetohydrodynamics (MHD) by Furth et al.~\cite{FurthKilleenRosenbluth_63}. The current sheet is unstable to large-scale tearing perturbations, which lead to magnetic reconnection and the formation of magnetic islands. In the nonlinear stage, an X-point configuration may collapse and form an elongated current sheet. In the Sweet--Parker model~\cite{Parker_57,Sweet_58a}, reconnection proceeds through such a current sheet. The current-sheet aspect ratio and reconnection rate are both controlled by the Lundquist number $S$, which is inversely proportional to the resistivity. The Lundquist number is typically very large in fusion and astrophysical plasmas.

At sufficiently large $S$, however, a long and thin current sheet is itself unstable to secondary tearing. Early theoretical and numerical studies of this instability were given by Bulanov et al.~\cite{BulanovSakaiSyrovatskii_79} and Biskamp~\cite{Biskamp_86}. Loureiro et al.~\cite{LoureiroSchekochihinCowley_07} later characterized the high-Lundquist-number instability of a Sweet--Parker current sheet, now commonly referred to as the plasmoid instability. Its disruption produces secondary plasmoids and additional current sheets, leading to a multiscale and intermittent reconnection process~\cite{BhattacharjeeHuangYang_09,ShibataTanuma_01}. Numerical simulations have shown that secondary plasmoids can move, merge, and interact with surrounding current sheets~\cite{DaughtonRoytershteynAlbright_09a,LuAngelopoulosArtemyev_20}. In gyrokinetic simulations, Numata and Loureiro~\cite{NumataLoureiro_15a} also observed the formation of a secondary plasmoid and its subsequent motion toward a primary island. Secondary magnetic islands have also been observed in space and astrophysical plasmas~\cite{EastwoodPhanMozer_07,Karlicky_04}.

Subsequent theories have considered the disruption of dynamically forming current sheets, including the effects of their temporal evolution~\cite{UzdenskyLoureiro_16, ComissoLingamHuang_16}. In the collisionless regime, Granier et al.~\cite{GranierBorgognoComisso_22, GranierNumataBorgogno_23} investigated current-sheet marginal stability using gyro-fluid and gyrokinetic approaches and demonstrated the importance of kinetic scales in determining the disruption threshold. These studies provide a framework for understanding the disruption of evolving current sheets that remain relatively uniform along their length. During the nonlinear evolution of reconnection, however, plasmoid motion and merger can generate substantial spatial nonuniformity and asymmetry along the current sheet. The role of such inhomogeneity in current-sheet disruption remains much less explored.

In this study, we investigate current-sheet disruption and secondary plasmoid formation using collisionless gyrokinetic simulations with the electromagnetic code AstroGK~\cite{NumataHowesTatsuno_10}. We compare simulations with and without small random perturbations in order to examine how symmetry breaking and the resulting spatial nonuniformity affect the subsequent nonlinear evolution. We quantitatively track X and O points, measure the widths and aspect ratios of current sheets immediately before disruption, and characterize the magnetic topology around secondary plasmoids.
These results provide a step toward understanding plasmoid instability in dynamically evolving and spatially inhomogeneous current sheets.

\section{Simulation}
\label{sec:simulation}

We perform simulations of magnetic reconnection under a strong guide magnetic field using the gyrokinetic code {\tt AstroGK}~\cite{NumataHowesTatsuno_10}. We consider a two-dimensional domain $-L_{x}/2 \leq x \leq L_{x}/2$, $-L_{y}/2 \leq y \leq L_{y}/2$ with periodic boundary conditions in both directions. The magnetic reconnection process is induced by the tearing instability, where a quasi-steady current sheet may be formed in the nonlinear stage. Depending on the parameters, the current sheets may undergo a secondary instability and generate plasmoids. We investigate numerically how nonuniformity along the current sheet affects the plasmoid generation.
All simulations are performed without collisions.

\subsection{Simulation setup}
\label{sec:setup}

The initial equilibrium magnetic field configuration $\bm{B}_0$ is given by 
\begin{align}
 \bm{B}_{0} = B_{z0} \zhat + B_{y0}^{\eq}(x) \yhat,
 \end{align}
where $B_{z0}$ is the out-of-plane guide magnetic field, $B_{y0}^{\eq} \ll B_{z0}$ is the in-plane reconnecting magnetic field. 
We set the reconnecting magnetic field as $B_{y0}^{\eq} = - \p A_{\parallel}^{\eq}/\p x$ with
\begin{align}
    A_{\parallel}^{\eq}(x)=A_{\parallel 0}^{\eq} \cosh^{-2} \left(\frac{x}{a}\right),
    \label{eq:apar_eq}
\end{align}
where the amplitude $A_{\parallel 0}^{\eq}$ and the scale length $a$ are constants. The initial reconnecting magnetic field defines the typical scale of the reconnection process. We use $a$ as the length scale and $\tau_{\mathrm{A}} = a/v_{\mathrm{A}}$ as the time scale, where $v_{\mathrm{A}}$ is the Alfv\'en velocity measured by the maximum of the initial magnetic field ($\max(B_{y0}^{\eq})\approx 0.77 A_{\parallel0}^{\mathrm{eq}}/a$).

The initial perturbation $A_{\parallel}^{\pt}$ is written as
\begin{align}
A_{\parallel}^{\pt}(y) = \sum_{k_y} A_{\parallel k_{y}} \cos(k_{y} y + \theta_{k_{y}})
\end{align}
where $A_{\parallel k_{y}}$, $k_{y}$, and $\theta_{k_{y}}$ are, respectively, the amplitude, wave number, and phase of the perturbation. We consider three types of initial perturbations: (i) imposing only the fundamental mode, $k_{y}^{(1)} = 2\pi/L_{y}$, (ii) imposing the second harmonic mode $k_{y}^{(2)} = 2k_{y}^{(1)}$, as well as the fundamental mode, (iii) imposing small random modes in addition to the type (i) or (ii) perturbations, with $A_{\parallel k_{y}}/A_{\parallel k_{y}}^{(1), (2)} \sim 10^{-4}$ and $\theta_{k_{y}}/(2\pi) \in [0, 1)$, where the phases are taken from a uniformly distributed random sequence. Note that $L_{y}$ may change depending on the cases. The box size in the $x$ direction is taken as sufficiently large $L_{x}/a = 6\pi$ so that the boundary does not affect the growth of magnetic islands.

\begin{figure}[htbp]
\centering
\includegraphics[scale=0.8]{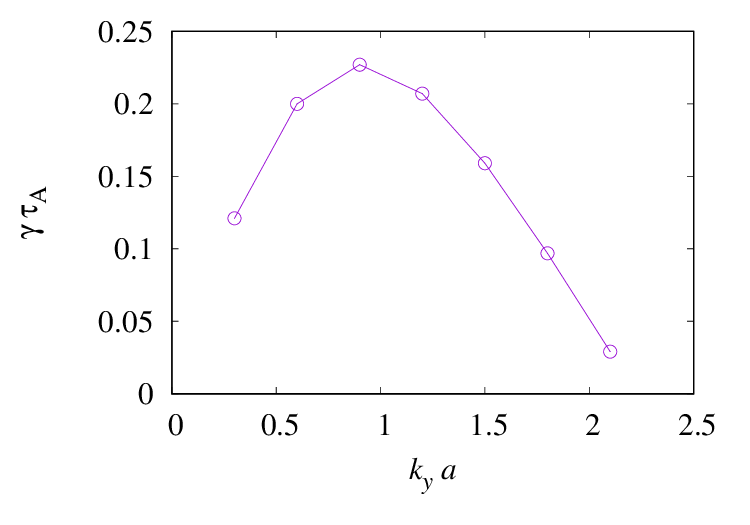}
\caption{~\label{fig:growth} Numerically calculated linear growth rate of the tearing mode using {\tt AstroGK}.}
\end{figure}

The choices of $k_{y}$ are summarized in Table~\ref{tbl:sim_cases}. The tearing mode stability parameter $\Delta'$ for the equilibrium \eqref{eq:apar_eq} is given by
\begin{align}
\Delta' a = 2 \left[ \frac{6 \tilde{k}_{y}^2 - 9}{\tilde{k}_{y} (\tilde{k}_{y}^2-4)} - \tilde{k}_{y} \right]
\end{align}
where $\tilde{k}_{y} \equiv \sqrt{a^2 k_{y}^{2} + 4}$. The mode numbers $k_{y}a = 0.3, 0.6$ are chosen to give large $\Delta'$ so that the linear growth rate (shown in Fig.~\ref{fig:growth}) is relatively large and plasmoids are known to be easily generated.
Other fixed physical parameters given in Table~\ref{tbl:phys_param} are the mass ratio, $m_{\mathrm{i}}/m_{\mathrm{e}}$, the temperature ratio, $T_{0\mathrm{i}}/T_{0\mathrm{e}}$, the electron plasma beta, $\beta_{\mathrm{e}} = 2\mu_{0} n_{0} T_{0\mathrm{e}}/B_{z0}^2$, and the ratio of the ion sound Larmor radius to the equilibrium magnetic field scale, $\rho_{\mathrm{Se}}/a$, where the ion sound Larmor radius is given by $\rho_{\mathrm{Se}}=\sqrt{m_{\mathrm{i}} T_{0\mathrm{e}}}/(e B_{z0})$, $m$ is the species mass, $T_{0}$, $n_{0}$ are the equilibrium temperature and density, respectively. We have set $n_{0\mathrm{i}}=n_{0\mathrm{e}}=n_{0}$ and consider singly charged ions.
The length scales relevant to the magnetic reconnection dynamics, also given in Table~\ref{tbl:phys_param}, are the Larmor radius and skin depth defined as
\begin{align}
\rho_{s} & = \frac{\sqrt{2m_{s} T_{0s}}}{e B_{z0}}
    = \rho_{\mathrm{Se}} \sqrt{2 \frac{m_{s}}{m_{\mathrm{i}}} \frac{T_{0s}}{T_{0\mathrm{e}}}}, &
d_{s} & = \sqrt{\frac{m_{s}}{\mu_{0} n_{0} e^{2}}} = \rho_{\mathrm{Se}} \sqrt{2 \frac{1}{\beta_{\mathrm{e}}} \frac{m_{s}}{m_{\mathrm{i}}}},
\end{align}
where $s=\mathrm{i}$ or $\mathrm{e}$ is the species index.

\begin{table}[htbp]
\centering
\caption{~\label{tbl:sim_cases} Parameters of initial perturbations for six simulation runs. The types of perturbations are (i) fundamental mode only, (ii) fundamental and second harmonic modes, (iii) small random perturbations in addition to (i) or (ii). The stability parameter $\Delta'a$ is for the fundamental mode.}
\vspace*{1em}
\begin{tabular}{ccccc|ccccc}
Case & Type & $k_{y}^{(1)} a$ & $k_{y}^{(2)}a$ & $\Delta'a$ &
Case & Type & $k_{y}^{(1)} a$ & $k_{y}^{(2)}a$ & $\Delta'a$ \\ \hline
1 & (i) & 0.3 & - & 167 & 4 & (iii) & 0.3 & - & 167 \\
2 & (i) & 0.6 & - & 41 & 5 & (iii) & 0.6 & - & 41 \\
3 & (ii) & 0.3 & 0.6 & 167 & 6 & (iii) & 0.3 & 0.6 & 167 \\
\end{tabular}
\end{table}

\begin{table}
\centering
\caption{~\label{tbl:phys_param} Physical parameters. The first four quantities are free parameters, which determine the ion/electron Larmor radii and skin depths.}
\vspace*{1em}
\begin{tabular}{lc}
Mass ratio & $m_{\mathrm{i}}/m_{\mathrm{e}} = 100$ \\
Temperature ratio & $T_{0\mathrm{i}}/T_{0\mathrm{e}} = 10^{-3}$ \\
Electron plasma & $\beta_{\mathrm{e}} = 0.2491$ \\
Ion sound Larmor radius & $\rho_{\mathrm{Se}}/a=0.3$ \\ \hline
Larmor radius of ion & $\rho_{\mathrm{i}}/a=0.013$ \\
Larmor radius of electron & $\rho_{\mathrm{e}}/a=0.042$ \\
Skin depth of ion & $d_{\mathrm{i}}/a=0.85$ \\
Skin depth of electron & $d_{\mathrm{e}}/a=0.085$
\end{tabular}
\end{table}

 \subsection{Overview of simulation results}
 \label{sec:results}

We give a summary of the simulation results. The system typically evolves as follows~\cite{LoureiroCowleyDorland_05}: After the initial linear tearing mode growth, the Rutherford stage where the magnetic islands grow algebraically follows. The X point in between the magnetic islands collapses to form a primary current sheet. Eventually, the primary current sheet also collapses due to the plasmoid instability, producing secondary plasmoids and secondary current sheets. The secondary current sheets may also follow the same process as the primary current sheet. Since the simulations are not externally driven, this sequence eventually ceases as the initial magnetic energy is depleted. However, we may expect the processes of current sheet formation and collapse to successively occur if sufficient magnetic energy remains.
During these processes, we observe the secondary plasmoids move and merge. 

We show some typical temporal evolution of the spatial distribution of the out-of-plane current $j_{\parallel}$ and exemplify how current sheets are generated and broken up by plasmoids.

\begin{figure}[htbp]
\centering
\includegraphics[scale=0.35]{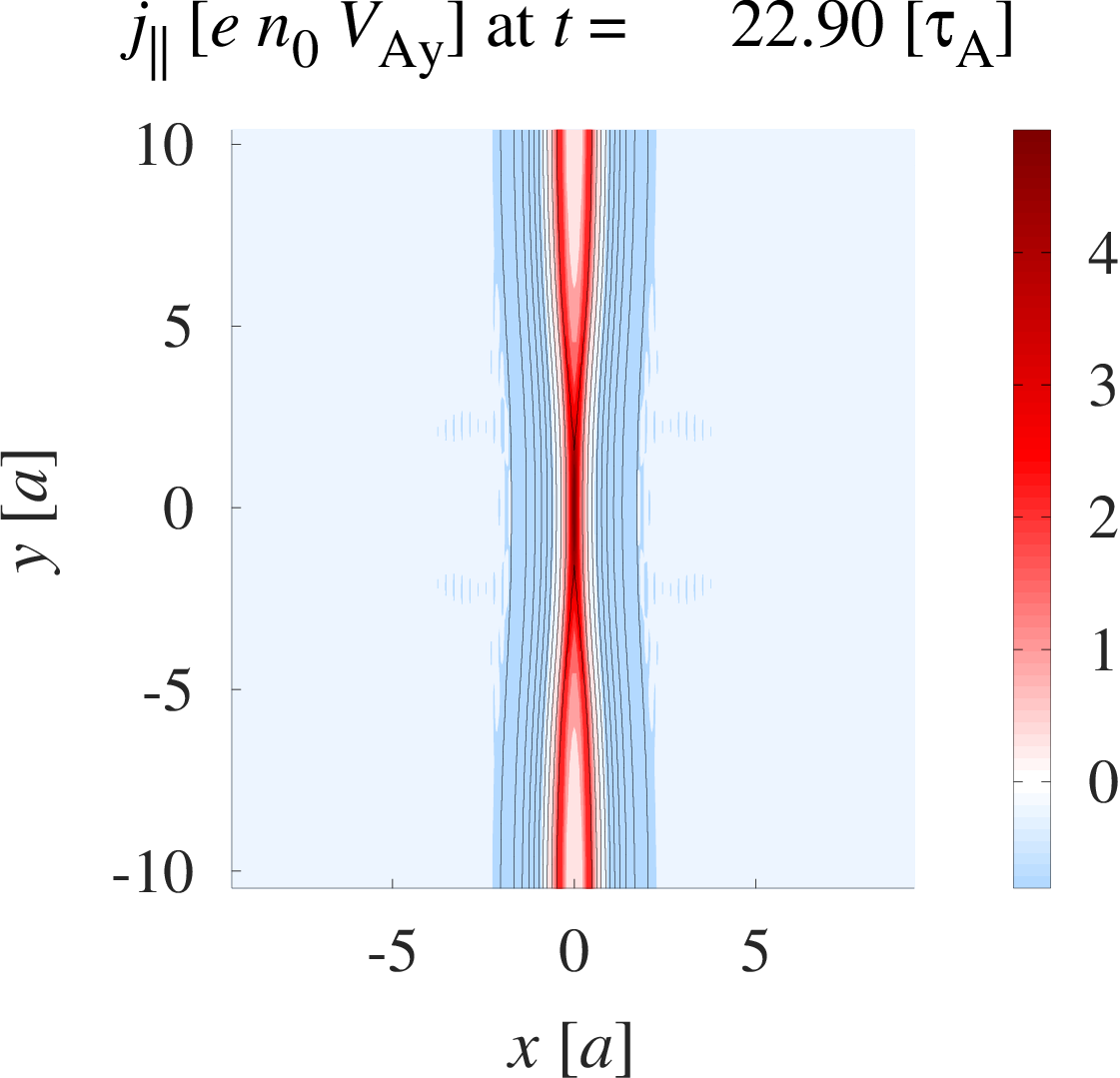}
\includegraphics[scale=0.35]{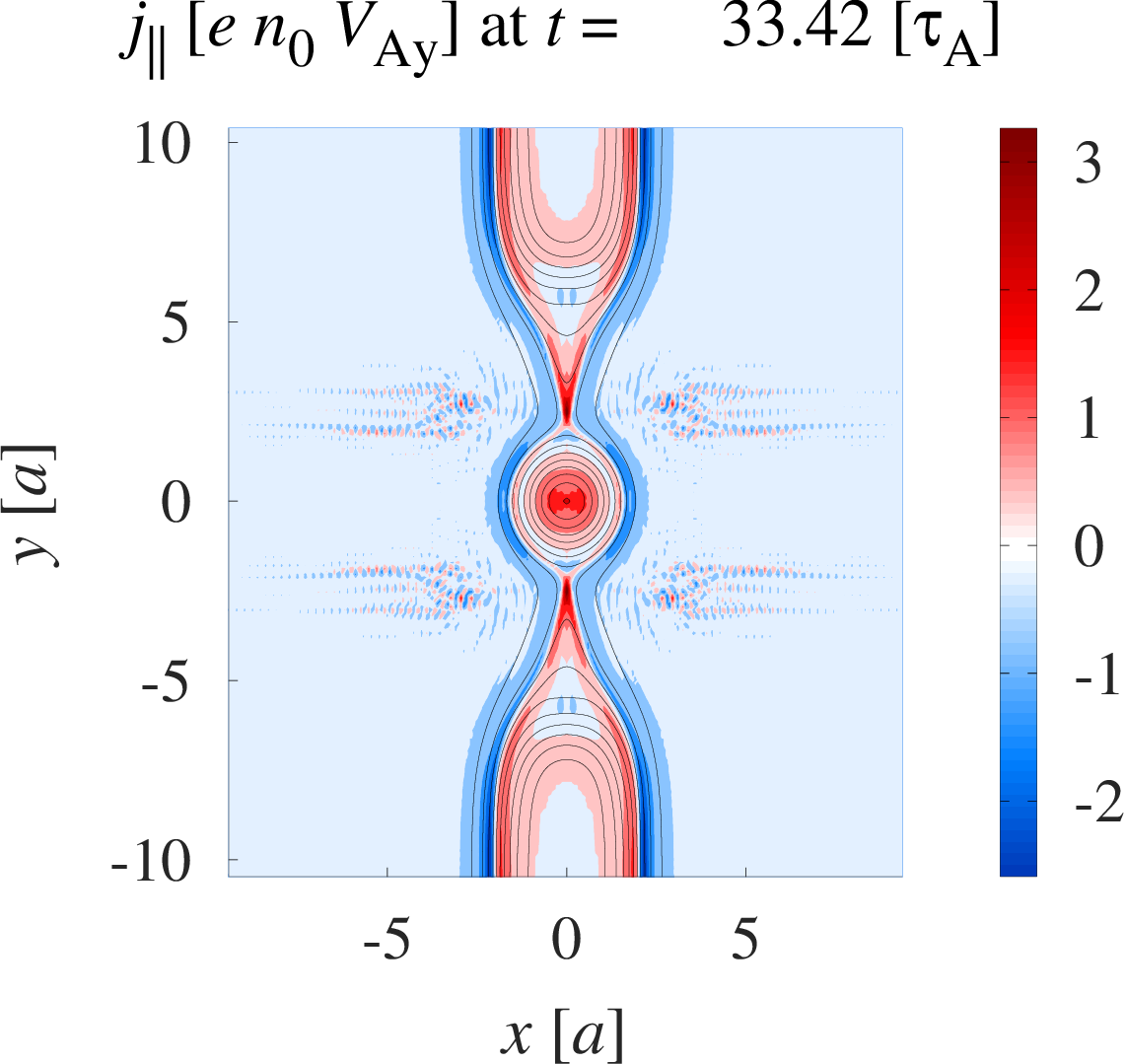}
\includegraphics[scale=0.35]{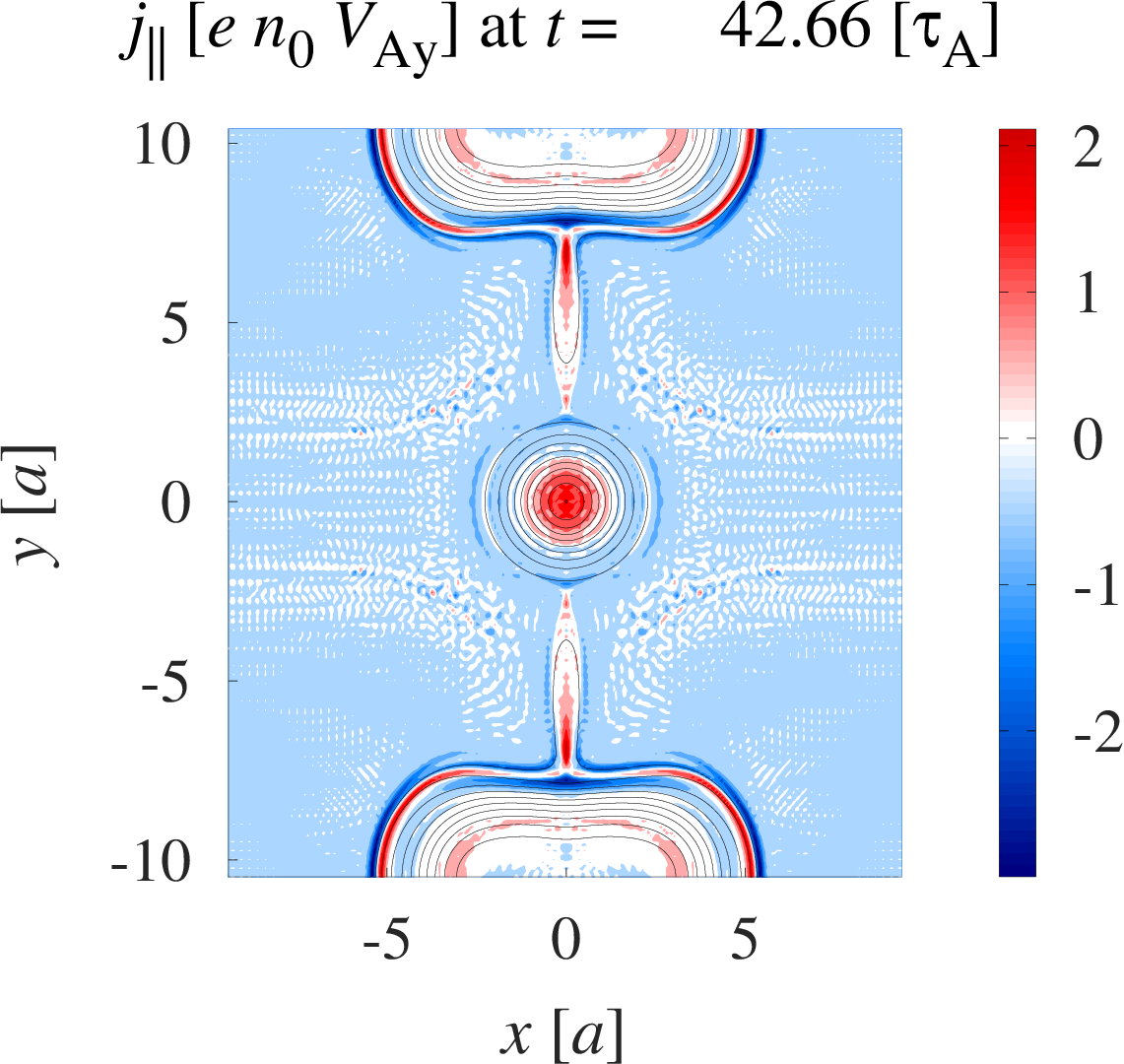}
\includegraphics[scale=0.35]{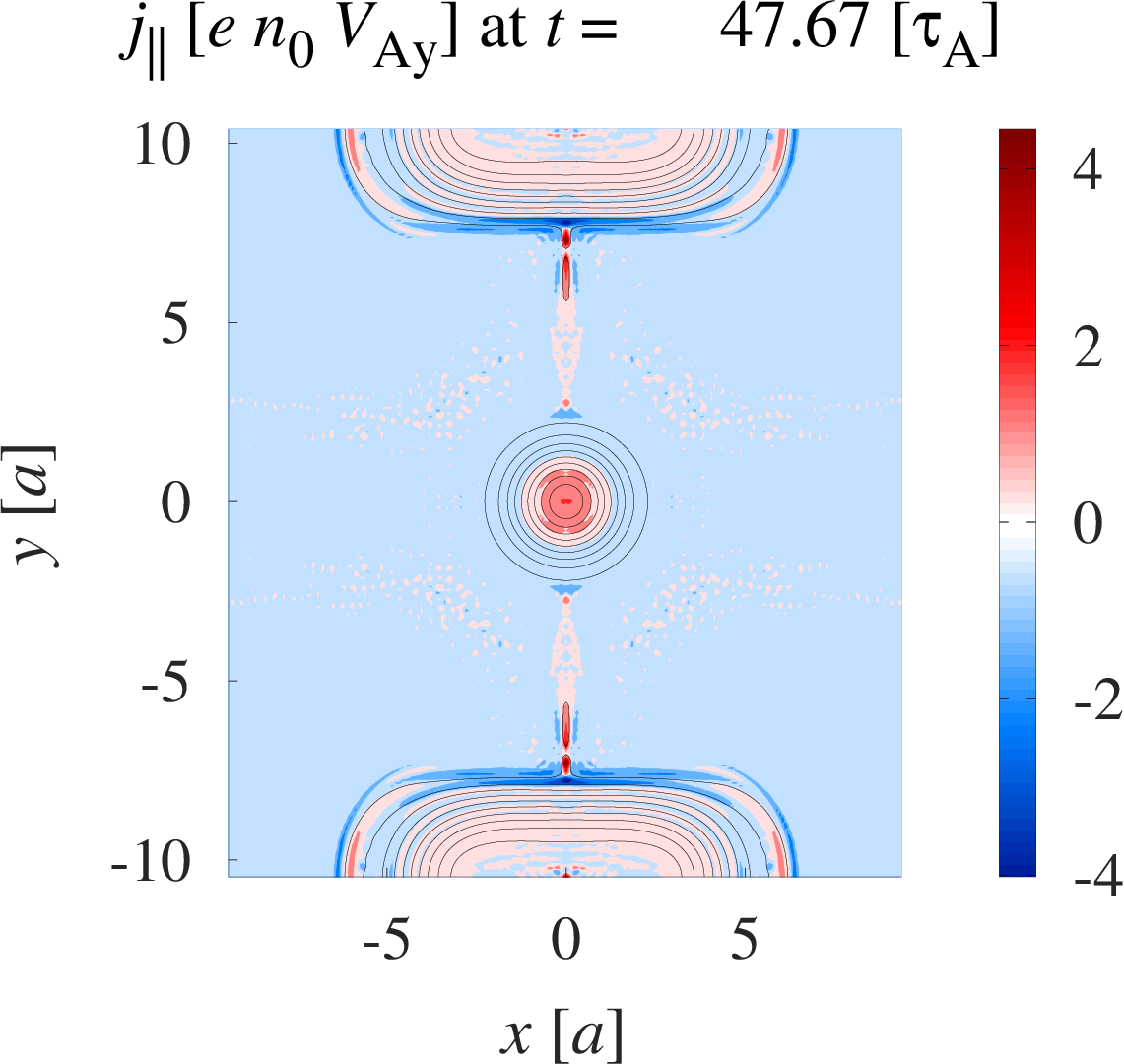}
\caption{~\label{fig:jpar_case1} Time evolution of the parallel current for Case 1. Snapshots are shown at $t/\tau_{\mathrm A}=22.90$, $33.42$, $42.66$, and $47.67$.}
\end{figure}

In Case 1 (Fig.~\ref{fig:jpar_case1}), the primary current sheet is formed during the Sweet--Parker-like stage of the nonlinear tearing mode evolution. The snapshot at $t/\tau_{\mathrm{A}}=22.9$ shows the elongated primary current sheet. It subsequently disrupts and produces a secondary plasmoid at the center of the domain ($y/a = 0$), as seen in the snapshot at $t/\tau_{\mathrm{A}}=33.42$. The secondary current sheets are newly formed in between the primary and secondary plasmoids. The secondary current sheets are pulled and are absorbed into the larger primary plasmoid. During the merger process, the secondary current sheets are compressed and are torn apart. New plasmoids are remnants of the torn current sheets.
The formation of the Y-shaped current sheet is also observed in~\cite{GranierBorgognoComisso_22}.
We also note here that the secondary plasmoid formed in the center of the domain stays for a long time because of the up-down symmetry of the simulation.

\begin{figure}[htbp]
\centering
\includegraphics[scale=0.35]{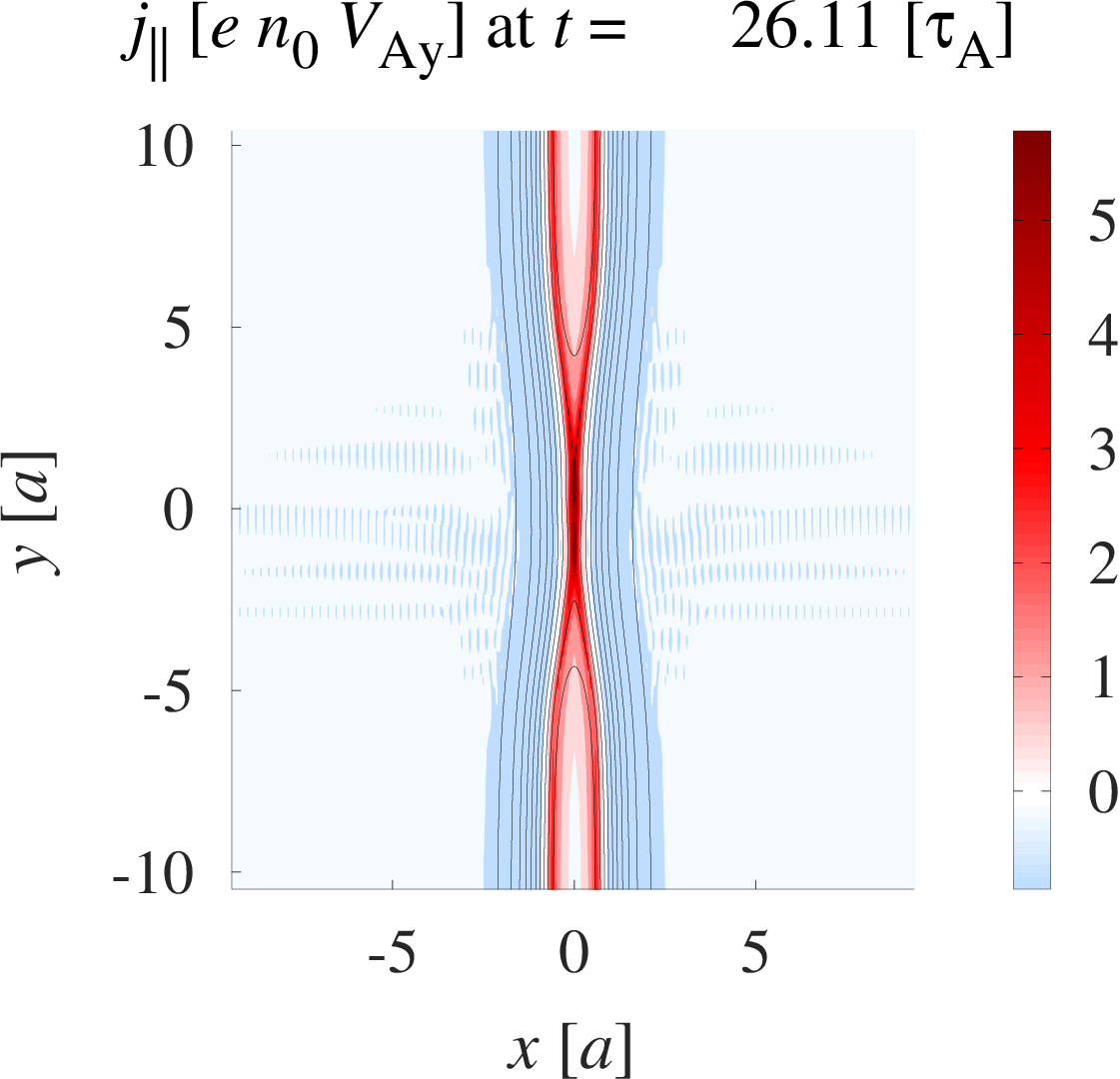}
\includegraphics[scale=0.35]{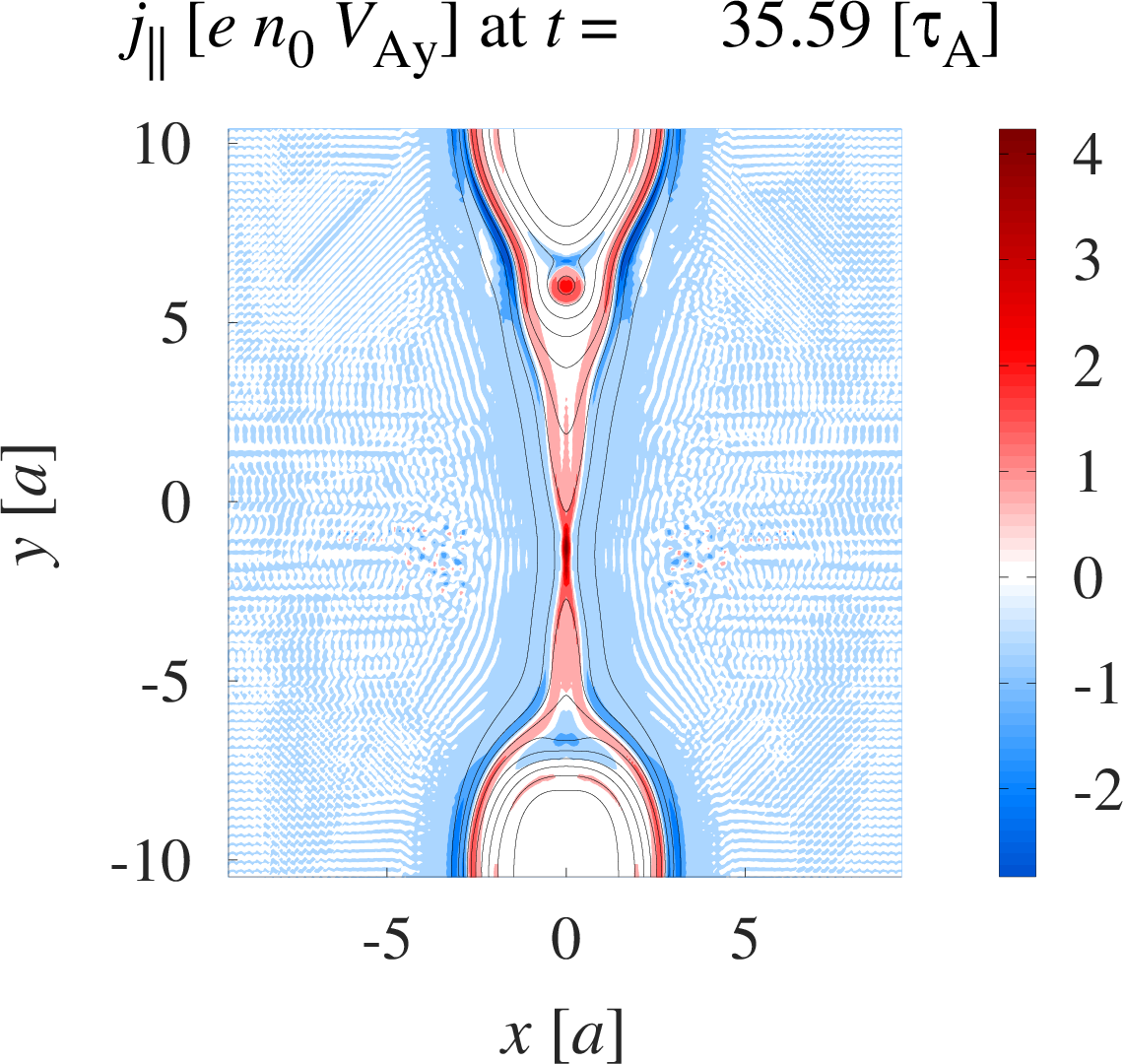}
\includegraphics[scale=0.35]{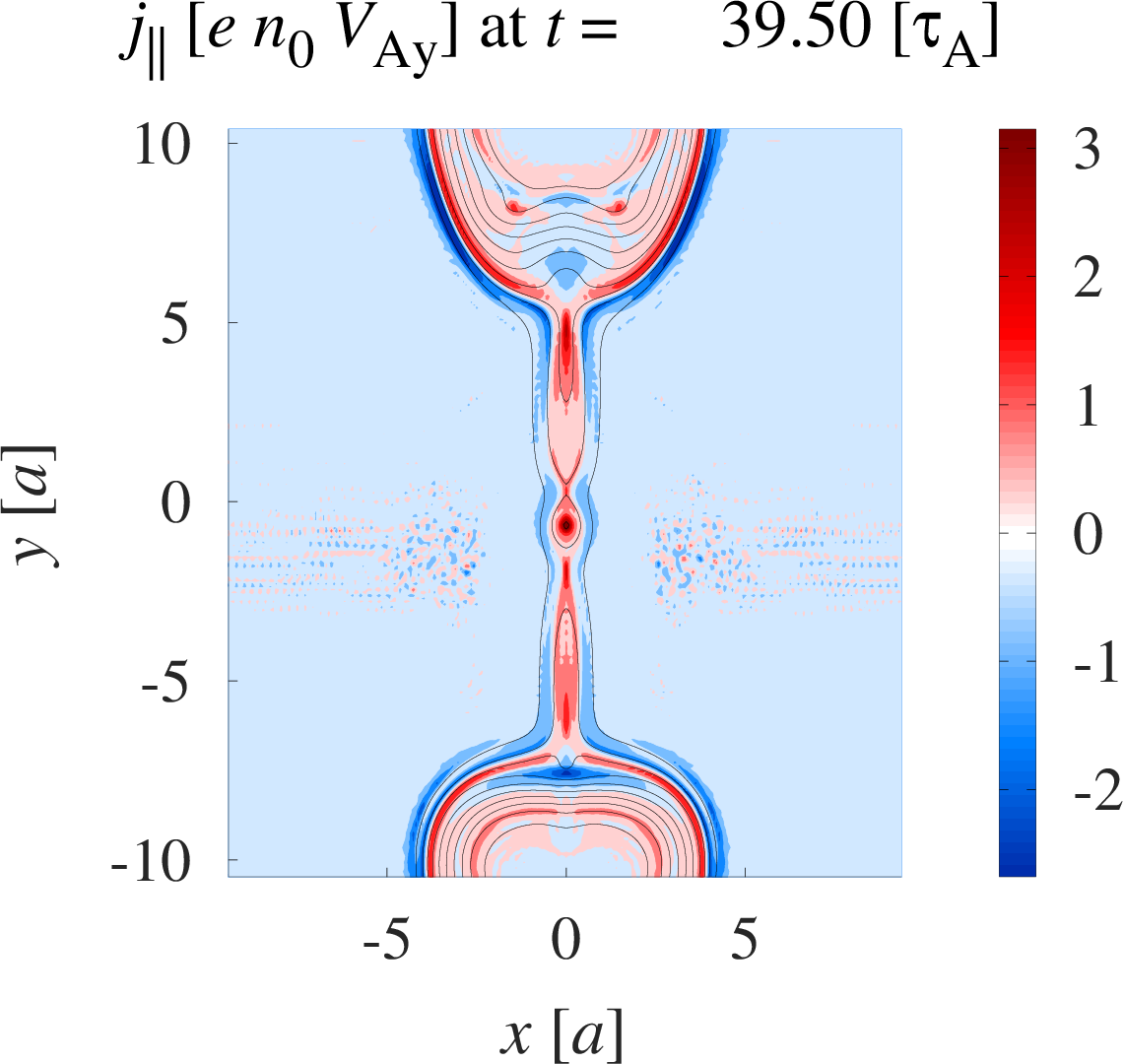}
\includegraphics[scale=0.35]{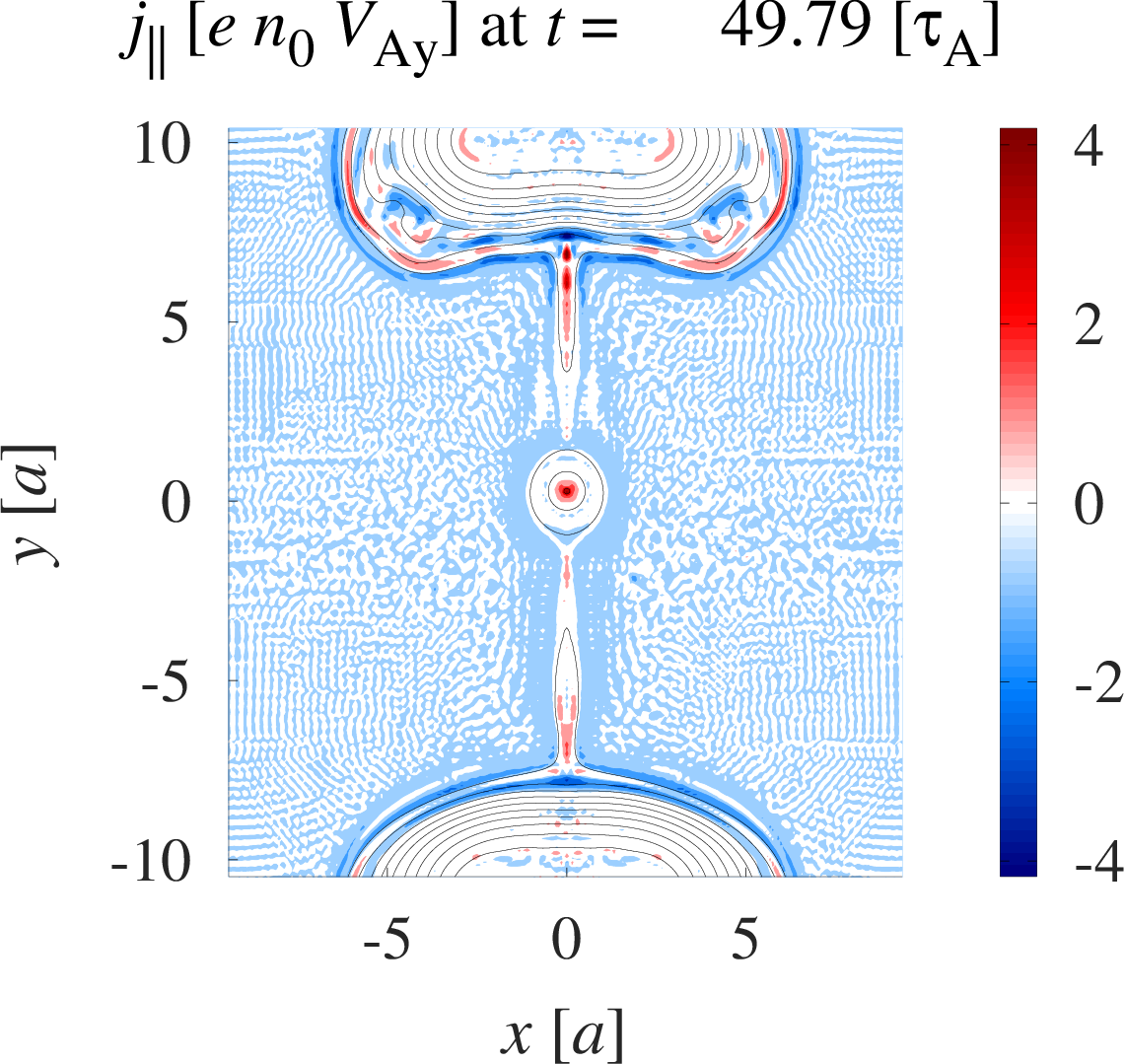}
\caption{~\label{fig:jpar_case4} Time evolution of the parallel current for Case 4. Snapshots are shown at $t/\tau_{\mathrm{A}}=26.11$, $35.59$, $39.50$, and $49.79$.}
\end{figure}

Random perturbations break the symmetry. Shown in Fig.~\ref{fig:jpar_case4} is the out-of-plane current for Case 4, where small random perturbations are superimposed on Case 1. Unlike Case 1, the secondary plasmoid in the center of the domain moves upward and coalesces into the primary island. When the secondary plasmoid moves out, it stretches the current sheet in the open site ($y/a \approx 0$). The same sequence of current sheet generation and break-up is repeated as long as sufficient magnetic energy remains. We also observe a similar current sheet generation and break-up sequence in the Y-shaped region where the plasmoid and current sheet are in contact.

\section{Properties of current sheets and plasmoids}
\label{sec:analyses}

The data obtained from the simulations are used to quantitatively detect and classify plasmoids and current sheets, and to evaluate their respective properties. Previous theories~\cite{GranierBorgognoComisso_22, ComissoLingamHuang_16, UzdenskyLoureiro_16} have shown that the aspect ratio is an important measure of current-sheet stability and plasmoid onset. We therefore measure the length $L_{CS}$ and width $d_{CS}$ of both primary and secondary current sheets and examine their aspect ratio at disruption. We also characterize the magnetic topology around the resulting plasmoids to investigate how random perturbations affect the subsequent evolution of current sheets and plasmoids.

\subsection{Current-sheet disruption and plasmoid dynamics}
\label{sec:classification}

\begin{figure}[htbp]
\centering
\includegraphics[scale=0.65]{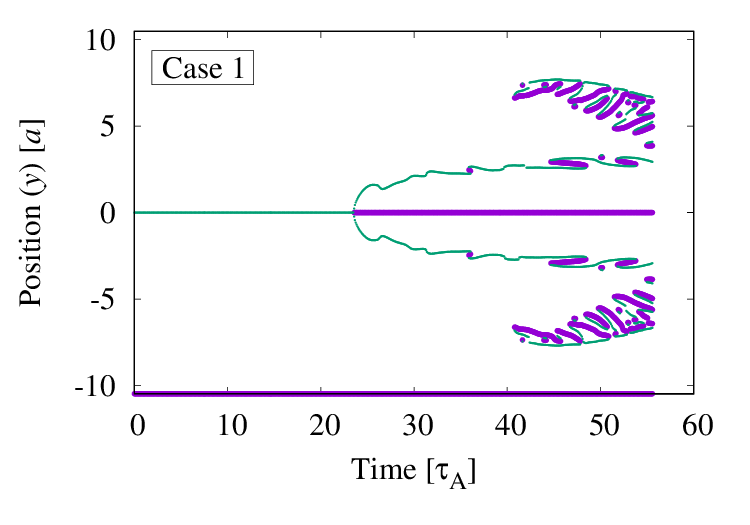}
\includegraphics[scale=0.65]{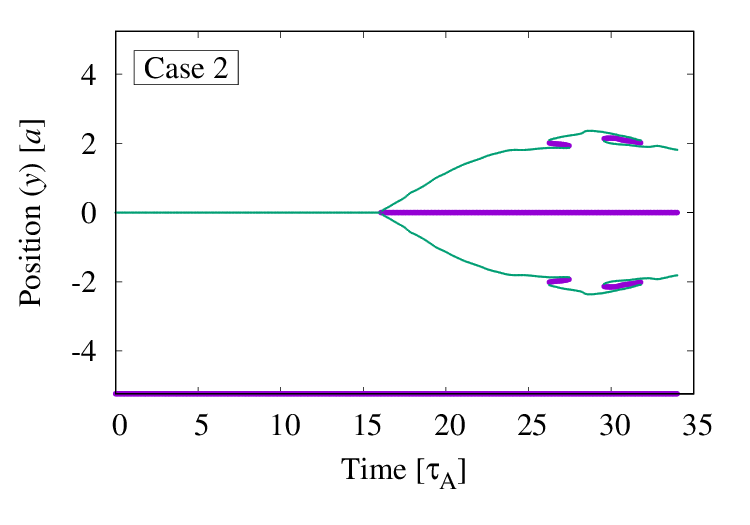}
\includegraphics[scale=0.65]{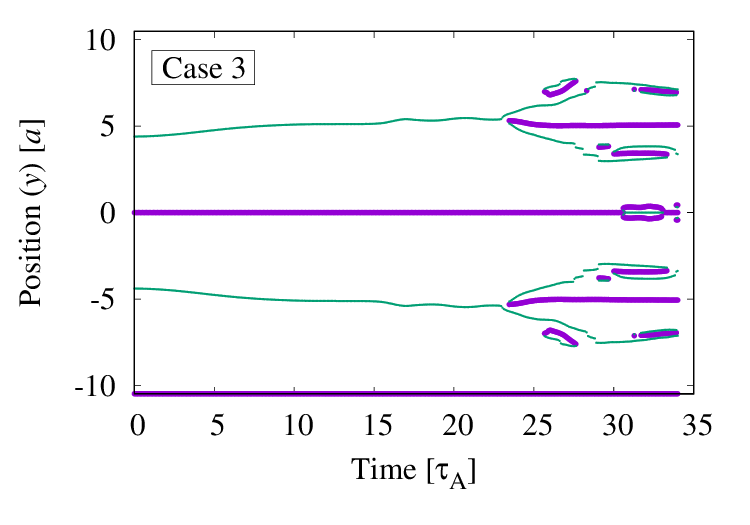}
\includegraphics[scale=0.65]{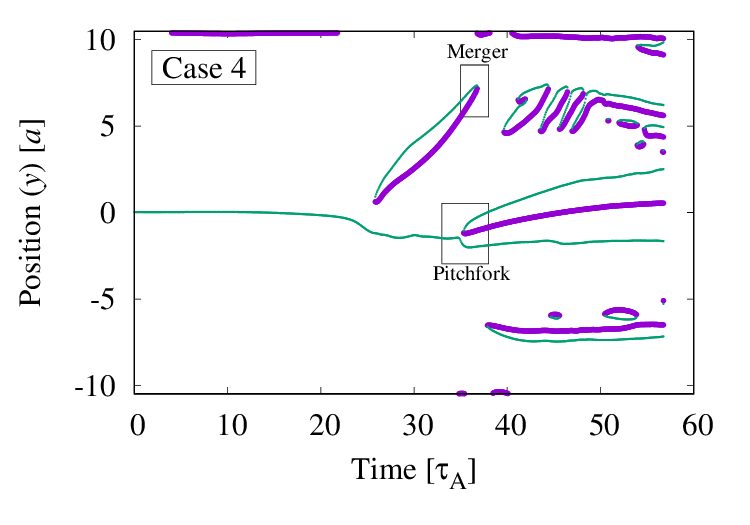}
\includegraphics[scale=0.65]{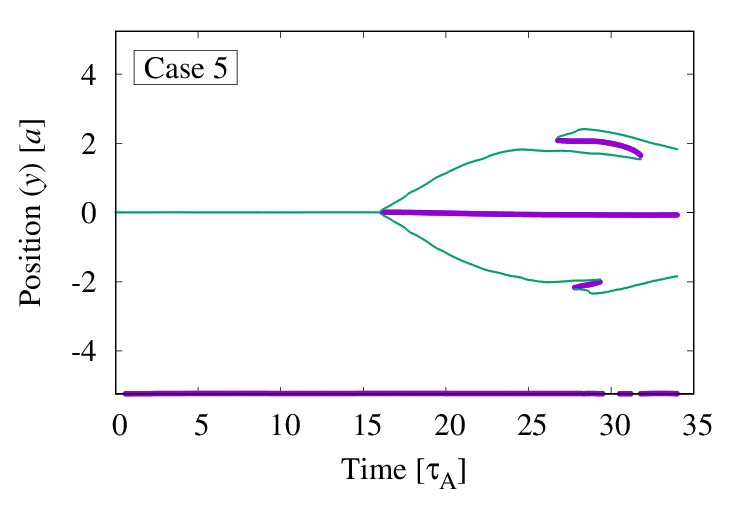}
\includegraphics[scale=0.65]{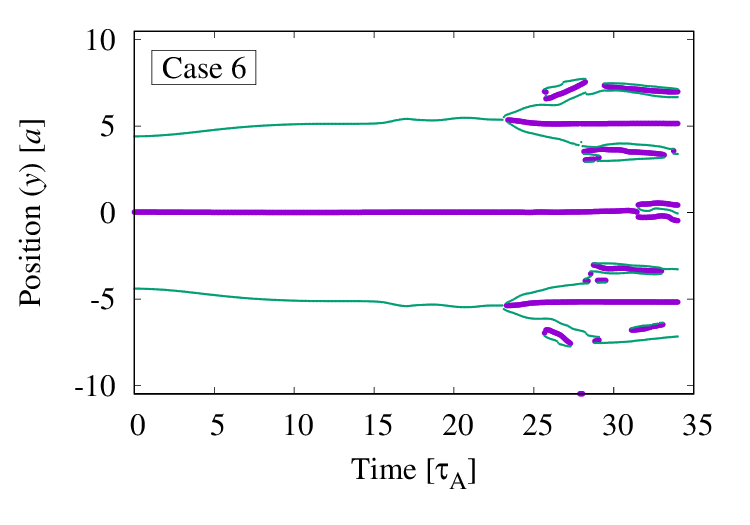}
\caption{~\label{fig:xo_trajectries} Trajectories of X and O points. The thin green lines/thick purple lines, respectively, denote X/O points.}
\end{figure}

To quantitatively identify current sheets and plasmoids, we examine the out-of-plane component of the vector potential $A_{\parallel}$. The current sheets and plasmoids are associated with the X and O points of the vector potential, respectively. For simplicity, we only consider the vector potential along the center line $x/L_{x}=0$, and look for the local minimum (X) and maximum (O) points. The X and O points always appear alternately. Figure~\ref{fig:xo_trajectries} shows the trajectories of X and O points by thin green and thick purple lines, respectively. Let us focus on Case 4 which includes all typical processes. Initially, there is one X point and one O point ($y/L_{y}\approx\pm 1/2$ is an O point). Around $t/\tau_{\mathrm{A}}\approx25$, an X point bifurcates like a pitchfork to form an O point in between X points.
The X point of the secondary current sheet generated by the preceding bifurcation also suffers a break-up like a pitchfork. These are typical plasmoid generations due to the break-up of a current sheet.
The square denoted by {\it Merger} shows the process of merger after the pitchfork. At $t/\tau_{\mathrm{A}}\approx25$, new X and O points appear, which finally merge at $t/\tau_{\mathrm{A}}\approx35$.
We also observe a series of generations and disappearances of pairs of X and O points during $t/\tau_{\mathrm{A}}=40\sim50$ around $y/a\approx 5$. These correspond to the events occurring in the Y-shaped region as described in the previous section.
Using these diagnostics, we can detect the time and location of the generation and disappearance of plasmoids and current sheets.

\begin{figure}[htbp]
\centering\includegraphics[scale=0.8]{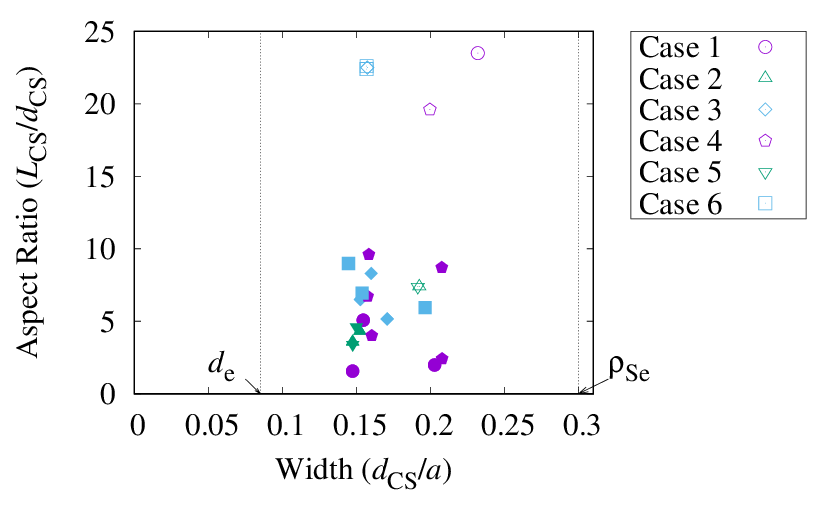}
\caption{~\label{fig:cs_width_length} Width and aspect ratio of the current sheets. Open symbols correspond to the primary current sheets, while filled symbols are for the secondary ones.}
\end{figure}

Previous studies have shown that the disruption threshold of an elongated current sheet can be characterized by its aspect ratio. In collisionless plasmas, the marginal stability also depends on kinetic scales such as the electron skin depth $d_{\mathrm{e}}$ and the ion-sound Larmor radius $\rho_{\mathrm{Se}}$~\cite{GranierBorgognoComisso_22}. Granier et al. showed that the marginal aspect ratio is not universal in the regime $\rho_{\mathrm{Se}} > d_{\mathrm{e}}$, which is relevant to the present simulations. We therefore measure the width $d_{\mathrm{CS}}$ and length $L_{\mathrm{CS}}$ of each current sheet immediately before disruption and define its aspect ratio as $A_{\mathrm{CS}}=L_{\mathrm{CS}}/d_{\mathrm{CS}}$.
Figure~\ref{fig:cs_width_length} shows the current-sheet width and aspect ratio immediately before disruption, which we refer to as the critical width and aspect ratio, respectively. The open symbols correspond to the break-up of primary current sheets.
The widths are confined to a relatively narrow range just above the electron skin depth, $d_{\mathrm{CS}}\gtrsim d_{\mathrm{e}}$, with little variation among the cases.
The primary current sheets reach substantially larger aspect ratios than the current sheets formed during the later nonlinear evolution.
The critical aspect ratios of the primary current sheets are of the same order as those expected from the gyro-fluid analysis of Granier et al.~\cite{GranierBorgognoComisso_22}.
The systematically smaller aspect ratios of the later current sheets motivate us to examine the magnetic topology surrounding them.

\subsection{Plasmoid topology and symmetry breaking}

\begin{figure}[htbp]
\centering
 \includegraphics[scale=0.4]{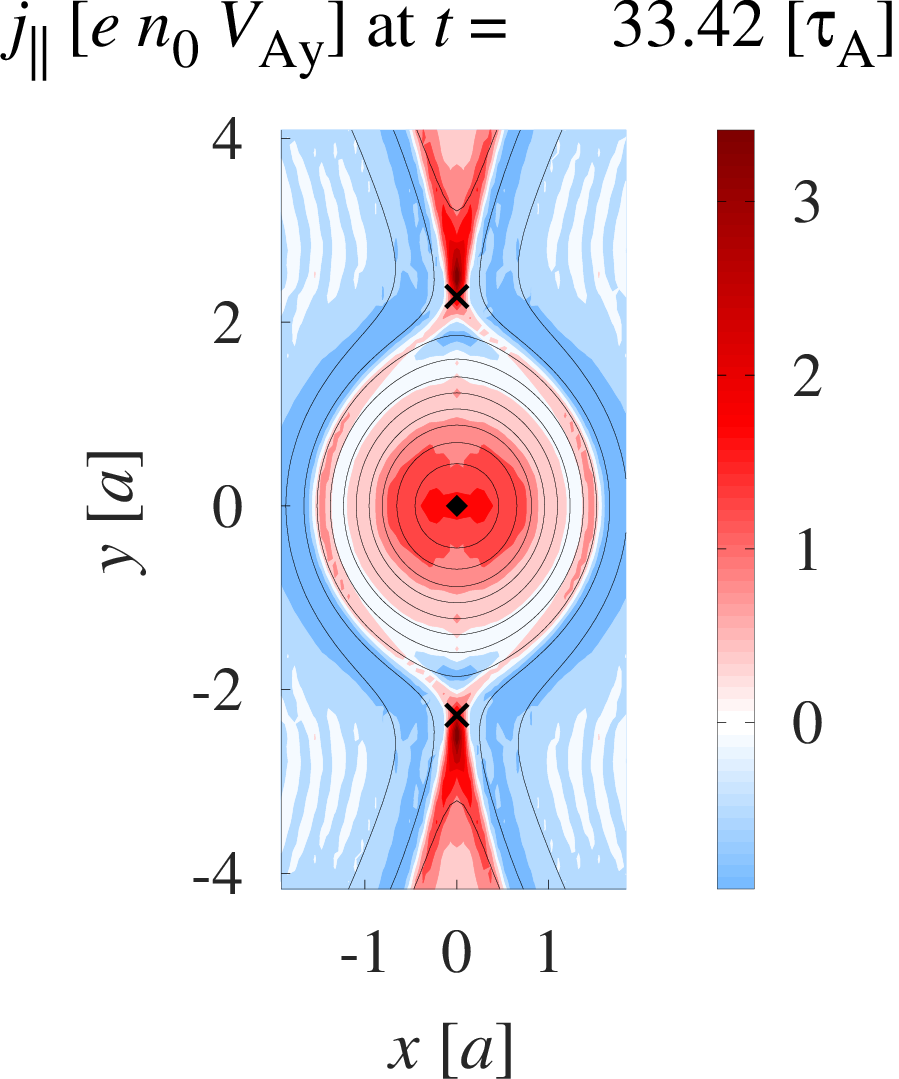}
 \includegraphics[scale=0.4]{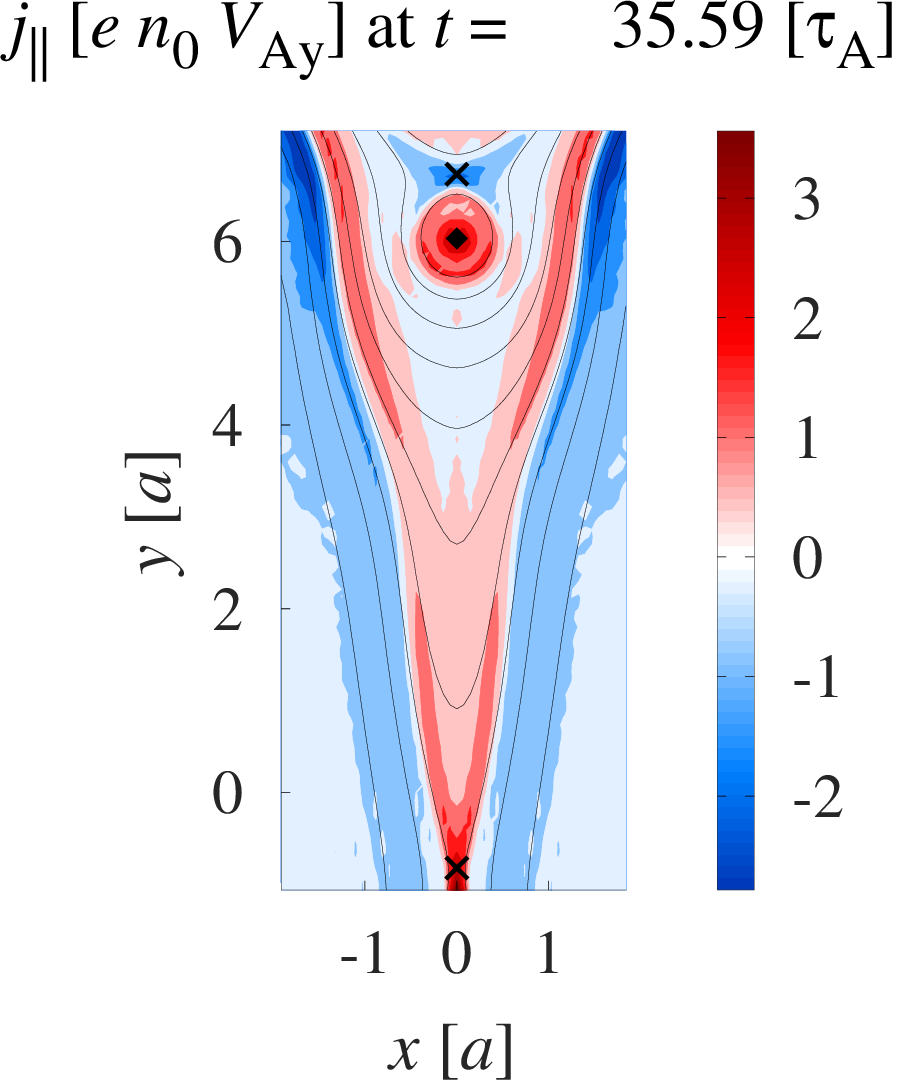}
\caption{~\label{fig:o_point_topology} Zoomed current-sheet profiles illustrate the topology around O points. The left panel shows the current profile at $t/\tau_{\mathrm{A}}=33.4$ for Case 1. The separatrix enclosing the O point passes through both neighboring X points, showing the cat's eye-like structure. The right panel shows the current profile at $t/\tau_{\mathrm{A}}=35.6$ for Case 4. The separatrix surrounding the O point at $y/a\approx6$ does not pass through the lower $X$ point. We refer to this type of magnetic structure around the O point as a teardrop topology.}
\end{figure}

A plasmoid is identified by an O point located between two neighboring X points.
The magnetic topology around a plasmoid can be classified according to whether the two neighboring X points lie on the same $A_{\parallel}$ contour. If they do, they are connected by a common separatrix.
We call the connected case a cat's-eye topology and the disconnected case a teardrop topology.
A teardrop topology typically appears in an asymmetric configuration, for example during plasmoid motion or merger.
Figure~\ref{fig:o_point_topology} shows examples of the two topologies observed in the simulations. In Case 1, the two neighboring X points lie on the same \(A_\parallel\) contour and form a common separatrix, giving a cat's-eye topology. In Case 4, the two X points have different \(A_\parallel\) values and are not connected by a common separatrix, giving a teardrop topology.

To quantify the difference between the two topologies, we measure the plasmoid width $w_{\mathrm{p}}$ using each of the two neighboring X points. For each X point, the plasmoid boundary is defined by the \(A_\parallel\) contour passing through that X point, and $w_{\mathrm{p}}$ is measured along the $x$ direction through the O point.

\begin{figure}[htbp]
\centering
 \includegraphics[scale=0.6]{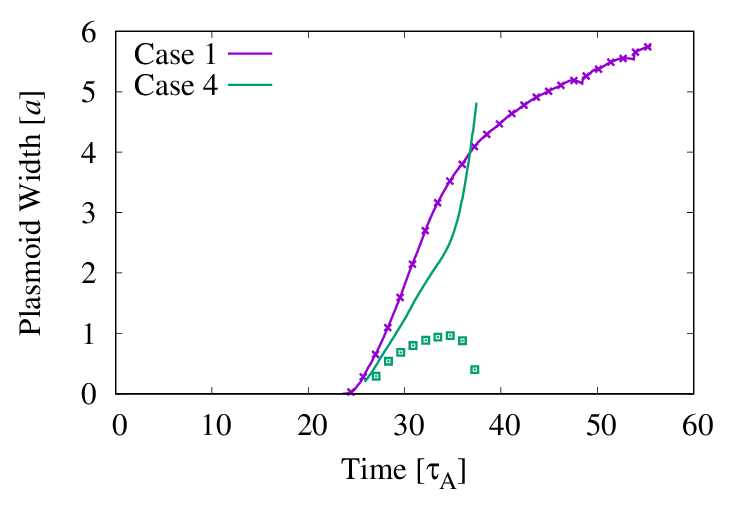}
\caption{~\label{fig:plasmoid_width} Plasmoid widths measured using the separatrices passing through the two neighboring X points. Solid lines denote the widths measured from the lower X point, while symbols denote those measured from the upper X point. For Case~1, the two widths are identical because the two X points lie on the same separatrix. For Case~4, the two widths differ because the neighboring X points belong to different separatrices.}
\end{figure}

Figure~\ref{fig:plasmoid_width} shows the temporal evolution of the plasmoid widths determined from the two neighboring X points. In Case 1, the two widths remain identical, consistent with the cat's-eye topology in which the two X points lie on the same \(A_\parallel\) contour. In Case 4, the two widths deviate from each other as the plasmoid evolves. Correspondingly, the values of \(A_\parallel\) at the two X points become different, and the common separatrix is lost.
Random perturbations break the symmetry and lead to an asymmetric, time-evolving magnetic topology around the plasmoid.
Such configurations illustrate how current sheets formed during the later nonlinear evolution can depart from the quasi-symmetric configurations considered in conventional stability analyses.

\section{Summary}
\label{sec:summary}
Magnetic reconnection simulations were performed using the electromagnetic gyrokinetic code AstroGK to investigate the formation and disruption of current sheets and plasmoids in collisionless plasmas. We considered six cases with different initial perturbations, including cases with small random perturbations that break the symmetry of the system. In all cases, the primary current sheet formed during the nonlinear evolution of the tearing mode eventually disrupted and generated secondary plasmoids and current sheets.

The X and O points were identified from the out-of-plane vector potential and tracked in time, allowing us to characterize the formation, motion, merger, and disappearance of plasmoids and current sheets. We measured the width and length of each current sheet immediately before disruption. The current-sheet widths were confined to a relatively narrow range just above the electron skin depth, whereas the disruption aspect ratios showed a clear difference between the primary current sheets and those formed during the later nonlinear evolution. The primary current sheets reached substantially larger aspect ratios, of the same order as those expected from the gyro-fluid analysis of Granier et al., while the later current sheets disrupted at smaller aspect ratios.

We also examined the magnetic topology around secondary plasmoids. Without random perturbations, the two neighboring X points can lie on the same separatrix, forming a cat's-eye topology. In the presence of random perturbations, the symmetry is broken and the two neighboring X points can have different values of \(A_\parallel\), producing a teardrop topology without a common separatrix. The corresponding plasmoid widths measured from the two neighboring X points also become different. These results show that random perturbations lead to asymmetric, time-evolving magnetic configurations around plasmoids, which differ from the quasi-symmetric configurations usually assumed in conventional current-sheet stability analyses.

These results suggest that the disruption of current sheets formed during the nonlinear evolution cannot always be understood solely in terms of the marginal stability of relatively uniform current sheets. The asymmetry and temporal evolution of the surrounding magnetic configuration may also play an important role in determining the disruption threshold. A systematic study varying the strength of the symmetry-breaking perturbations and the relevant kinetic scales, together with statistics over a larger number of current sheets, will be required to quantify this effect and to establish a stability criterion applicable to dynamically evolving current sheets.

\section*{Acknowledgement}
The computation of this work is performed on the facilities of the Center for Cooperative Work on Computational Science, University of Hyogo, the JFRS-1 supercomputer system at the Computational Simulation Centre of the International Fusion Energy Research Centre (IFERC-CSC) in Rokkasho Fusion Institute of QST (Aomori, Japan) and on the ‘Plasma Simulator’ of NIFS with the support and under the auspices of the NIFS Collaboration Research program (NIFS18KNSS112).

\bibliographystyle{unsrt}
\bibliography{asym_plasmoid}

\end{document}